\documentclass[useAMS,usegraphicx,usenatbib]{mn2e}
\usepackage{epsfig}
\usepackage{color}

\usepackage{amssymb}
\usepackage{amsmath}

\newcommand{\be}{\begin{equation}}
\newcommand{\ee}{\end{equation}}

\newcommand{\msunh}{${\rm M}_{\odot} h^{-1}$\,}

\newcommand{\ns}{${\rm n}_{s}$\,}
\newcommand{\omb}{$\Omega_{b}$\,}
\newcommand{\omdm}{$\Omega_{m}$\,}
\newcommand{\oml}{$\Omega_{\Lambda}$\,}

\newcommand{\nsigm}{$\sigma_{8}$\,}

\title[Spin distributions of LCDM haloes]{Dissecting the spin distribution of Dark Matter haloes}
\author[V. Antonuccio-Delogu et al.]{\noindent
  V.~Antonuccio-Delogu$^{1,3, 4}$\thanks{E-mail: Vincenzo.Antonuccio@oact.inaf.it}, A.~Dobrotka$^{2}$\thanks{E-mail: andrej.dobrotka@stuba.sk}, U.~Becciani$^{1}$, 
  S.~Cielo$^{1,4}$, C.~Giocoli$^{3}$, \and A. V. Macci\`{o}$^{5}$ and A. Romeo-Velon\'{a}$^{6}$
\\~\\
$^1$ INAF -- Osservatorio Astrofisico di Catania, Via S. Sofia 78,
  Catania, I-95123, ITALY \\
$^2$ {Department of Physics, Institute of Materials Science, Faculty of Materials Science and Technology, Slovak University}\\
\noindent
{of Technology in Bratislava,  J\'{a}na Bottu 25, 91724 Trnava, The Slovak Republic}\\
$^3$ Institut f\"{u}r Theoretische Astrophysik, Ruprechts-Karls-Universit\"{a}t, Albert-Ueberle-Stra\ss e 2, 69120 Heidelberg, Germany\\
$^4$ Scuola Superiore di Catania, Via San Nullo, 5/i, 95123 Catania, Italy\\
$^5$ Max-Planck-Institut f\"{u}r Astronomie, Konigst\"{u}hl 17, D-69117 Heidelberg, Germany\\
$^6$ Universidad Andres Bello, Departamento Ciencias Fisicas, Av. Republica
220, Santiago, Chile}

\begin{document}

\date{Accepted ??. Received ??; in original form 2007 ??}

\pagerange{\pageref{firstpage}--\pageref{lastpage}} \pubyear{2007}

\maketitle


\label{firstpage}

\begin{abstract}
The spin probability distribution of Dark Matter haloes has often been modelled as being very near to a lognormal. Most of the theoretical attempts to explain its origin and evolution invoke some hypotheses concerning the influence of tidal interactions or merging on haloes. Here we apply a very general statistical theorem introduced by Cram\'{e}r (1936) to study the origin of the deviations from the reference lognormal shape: we find that these deviations originate from correlations between two quantities entering the definition of spin, namely the ratio $J/M^{5/2}$ (which depends only on mass) and the modulus $E$ of the total (gravitational + kinetic) energy.\\
\noindent
To reach this conclusion, we have made usage of the results deduced from two high spatial- and mass resolution simulations. Our simulations cover a relatively small volume and produce a sample of more than 16,000 gravitationally bound haloes, each traced by at least 300 particles. We verify that our results are stable to different systematics, by comparing our results with those derived by the \textsc{GIF2} and by a more recent simulation performed by Macci\`{o} et al.\\
We find that the spin probability distribution function shows systematic deviations from a lognormal, at all redshifts $z \la 1$. These deviations depend on mass and redshift: at small masses they change little with redshift, and also the best lognormal fits are more stable. The $J-M$ relationship is well described by a power law of exponent $\alpha$ very near to the linear theory prediction ($\alpha=5/3$), but systematically lower than this at z$\la 0.3$. We argue that the fact that deviations from a lognormal PDF are present only for high-spin haloes could point to a role of large-scale tidal fields in the evolution of the spin PDF.
\end{abstract}


\section{Introduction}
Within the hierarchical galaxy formation model, Dark Matter (hereafter DM) haloes are thought to play the role of gravitational building blocks, within which baryonic diffuse matter collapses and becomes detectable. On galactic scales, the formation of stars and their evolution provides an important probe of the evolution of the visible content of the Universe \citep{1978MNRAS.183..341W,1991ApJ...379...52W}, although the subtleties of the stellar formation processes within galaxies, as of today not yet completely understood, hinders an exploitation of these objects as a clean probe of the evolution of DM haloes. On the other extreme of the mass scale, the most massive clusters of galaxies are regarded as one of the most reliable cosmological probes \citep{1997ApJ...485L..53B}: in particular, their abundance and evolution with redshift is a very sensitive test of the underlying cosmological model \citep{1993Natur.366..429W,1997ApJ...489L...1H,1997ApJ...485L..53B,1997A&A...317....1O,1998MNRAS.298.1145E,1999ApJ...523L.137D,2000ApJ...534..565H,2002A&A...386..775A,2004ApJ...609..603H,2007ApJ...655..128G}.\\
\noindent
The Mass Function (hereafter MF) is an indirect test of the total virialised mass of DM haloes: exact predictions of the latter can be done using the nonlinear spherical collapse model \citep{1972ApJ...176....1G}, an essential ingredient of the Press-Schechter model for the MF. However, DM haloes also possess angular momentum, and taking into account its effect on the gravitational collapse of rotating shells has been shown to have a detectable consequence on  the MF \citep{2006A&A...454...17D,2009ApJ...698.2093D}. Thus, an investigation of the distribution of angular momentum, and its connection with mass is particularly useful, as high-resolution simulations of increasing resolution will produce MFs with a very small statistical uncertainty.\\
\noindent 
An exact determination of the shape of the DM halo spin PDF can also have important consequences for the abundance of \emph{low surface brightness galaxies}, if the latter form preferentially within high-spin DM haloes \citep{1998MNRAS.299..123J, 2007MNRAS.378...55M}. Finally, it 
is of great importance also in models of the formation and evolution of central Black Holes. In collapse models where most of the baryons' specific angular momentum is a fixed fraction of that of their host dark matter haloes \citep{1980MNRAS.193..189F,1998MNRAS.295..319M}, the angular momentum and extent of the gaseous central accretion disc are strongly dependent on their's spin. \citet{2005ApJ...633..624V} find that the central density of the disc varies as $\rho_{0} \simeq\lambda^{-4}$, thus the initial rate of accretion of the central Black Hole turns out to be a very sensitive function of the spin $\lambda$ \citep{1971A&A....11..377P}.\\
\noindent For all these reasons, investigations of the origin of the angular momentum growth and of the spin PDF of DM haloes, and of their evolution, can have an impact on many different open issues in large-scale structure formation and evolution.
\\
\noindent Theoretical investigations \citep{1969ApJ...155..393P} predict that to zeroth-order the angular momentum should have a power-law dependency on the total virialized mass, with exponent $5/3$. They also make a prediction concerning the \emph{spin}, a dimensionless quantity defined as:
\be
\lambda = \frac{J E^{1/2}}{GM^{5/2}}
\label{eq:am:0}
\ee
where $J\equiv\mid\bmath{J}\mid$, $E\equiv\mid E_{kin} + E_{grav} \mid$ are respectively the moduli of the angular momentum, and of the total (kinetic plus potential) energy. The PDF of $\lambda$ has been predicted to have an approximately lognormal distribution. This initial prediction was subsequently found to be basically valid also when higher order effects were taken into account \citep{1996MNRAS.282..436C,1996MNRAS.282..455C}. More recently, the radial profile of $\lambda$ has been derived from modified Jeans' equations \citep{2009ApJ...694..893S}, and found to be in good agreement with results from simulations.\\
\noindent
The properties of the spin distribution of DM haloes have recently been considered, taking advantage of the availability of high spatial- and mass-resolution simulations. \citet{2007MNRAS.376..215B} have analysed a large sample of haloes drawn from the Millennium simulation, and found (among other things) that the \emph{global} spin distribution is poorly described by a lognormal distribution:
\be
P(\lambda) = \frac{1}{\lambda\sqrt{2\pi\sigma_{\log{\lambda}}^{2}}}\exp\left(-\frac{{\rmn \log}^{2}(\lambda/\bar\lambda)}{2\sigma_{\log{\lambda}}^{2}}\right) \label{eq:1}
\ee
 They alternatively suggest an empirical fit:
\be
P(\lambda) \propto \left(\frac{\lambda}{\lambda_{0}}\right)^{3}\exp\left[-\alpha\left(\frac{\lambda}{\lambda_{0}}\right)^{3/\alpha}\right]
\ee
which has the shape of a lognormal except at very high and very low values of the spin. They also suggest that the actual shape of the distribution depends on the adopted \emph{numerical definition} of halo.\\
\noindent
Theoretical and numerical studies are aimed at understanding the
origin of the (almost) lognormal spin distribution, using a
limited set of statistical and dynamical assumptions or by performing
\emph{controlled} numerical experiments. \citet{2002ApJ...581..794C} have shown that the distribution of specific angular momentum (which they define as $j' = J/M^{5/3}$) can be described by a rather complex PDF, which can be approximated by a lognormal in the central part, but deviates significantly from it at low- and high values of $j'$. They also make predictions for the dependence of the peak of the spin distribution for different values of mass, suggesting that it varies linearly with the average $\sigma_{\delta}$ of the distribution. \citet{2008ApJ...678..621K} find a correlation of spin $\lambda$ with mass, albeit a weak one. Note that they use a set of simulations using boxes of varying sizes, while in the present work we adopt a single simulation, thus minimising the spurious tidal effects.
Furthermore, \citep{2007MNRAS.376..215B,2007MNRAS.377L...5G,2008MNRAS.389.1419L}, studying the spin distributions at low redshifts, have found that more massive haloes show larger values of $\lambda$. Extensions of these results to higher redshift have been provided by \citet{2009MNRAS.393.1498D}, who have studied the evolution of the spin
distributions at $z > 6$, and showed that more massive haloes ($M\simeq
10^{7} {\rm M}_{\odot}$) tend to have a median $\lambda$ higher than that of $M\simeq
10^{6} {\rm M}_{\odot}$ haloes. 
Also, high-$\lambda$ haloes tend to
cluster more (by a factor 3-5) than low-spin haloes, a trend which
strengthens with time. However, their simulation is restricted to a
small box ($L_{b} = 2.46 h^{-1} {\rm Mpc}$), thus making their result more
prone to uncertainties from cosmic variance.\\
\noindent
The steady improvement of the available hardware and software resources makes today possible simulations where the limits imposed by the finite spatial and mass resolution limits are challenged. The simulations we have performed in this work were aimed at providing a \emph{statistically significant} sample of reasonably well-resolved DM haloes. We have obtained a catalogue of more than 77600 DM haloes, each resolved by more than 20 particles, and we have used only those haloes with more than 300 particles, thus resulting in a catalogue containing more than 16400 haloes. We have chosen a box size of $L = 70 \, h^{-1}$ Mpc, smaller than the one used in previous papers \citep[e.g.][]{2006ApJ...646..815S}, and a large number of particles, to maximize the mass resolution.\\
\noindent
In this work we present the results of a numerical experiment, where we study the evolution of the angular momentum/spin distributions within a relatively small cosmological volume. We find that the spin distribution shows some clear deviations from a lognormal, and does not seem to relax towards the latter, above statistical uncertainty. These deviations \emph{do not seem to depend significantly on the redshift or the mass of the halo}, at least for the underlying LCDM model studied in this paper. We also perform a possible check on the role of possible systematics in our analysis by comparing with the results of two different simulations: the GIF2 \citep{2008MNRAS.386.2135G}, and a simulation recently perfomed by \citet{2008MNRAS.391.1940M}. Both these works adopted a Spherical Overdensity (SO) \emph{halo finder}, slightly different from our chosen halo finder (the Amiga Halo Finder, \textsc{AHF}), recently introduced by \citet{2009ApJS..182..608K}. We address the problem of the origin of these deviations by exploiting an exact result from statistics \citep[\emph{Cram\'{e}r's theorem},][]{Cramer..MatZeit..41..405..1936} to demonstrate that the deviations from a lognormal distribution are induced from \emph{correlations} between the total energy and mass of haloes, as should be expected for not completely virialised haloes.\\
\noindent
The plan of the paper is as follows: In section~\ref{sec:sims} we present the details of the simulations we have performed, briefly describing the code adopted, the initial conditions and the halo finder. We then describe the main results concerning $J-M$ relationship (section~\ref{sec:jm}). In section~\ref{sec:diss:spin} we concentrate on the spin probability distribution, and we show that the deviations from a lognormal shape are dependent on halo mass. We trace the origin of these deviations to the presence of \emph{detectable} correlations among the quantities entering its definition, by applying the Kendall test. We discuss our findings in the Conclusion (section~\ref{sec:concl}).\\
In the following, we will denote the natural logarithm using the symbol ``$\ln$'', and the decimal logarithm as ``$\log$'' (or ``Log'').
  
\section{Numerical simulations} \label{sec:sims}
As underlying cosmological model we have chosen the fiducial
5-year WMAP LCDM cosmology \citep{2009ApJS..180..306D}, with Hubble constant: $H_{0} = 71.9\,$, and the cosmological parameters: 
(\omb, \oml, \omdm, \ns, \nsigm) = $[0.044, 0.742, 0.258, 0.963, 0.7986]$.\\
\noindent
Our main target was that of producing a statistically significant number of DM haloes, thus we tried to adopt a softening length sufficiently small to ensure that the smallest haloes we want to resolve have a size large enough at the initial redshift of the simulation. In order to do this, we define at first an initial instability radius $r_{initial}$, for which we refer to the ${\rm r}_{200}$ radius presented in eq. (1) of \citet{2002MNRAS.336..112M}: $r_{200}~= 2.98\times 10^{2}(M_{9}/\Omega_{m})^{1/3}/(1+z)\,$ kpc (where $M_{9}$ is defined as $M/10^{9}M_{\odot}$). For the initial redshift of our simulations, their formula gives: $r_{initial}=r_{200}(z=50,M_{m}) \approx 10.36\,$ kpc. Moreover, as we explain later, we consider only haloes having a lower mass threshold $M_{m} \geq 1.44\times 10^{9} \rm{M}_{\sun}$. We then choose a softening length $l = 12.5\,$ kpc, thus allowing the linear density field to become gravitationally unstable on this spatial scale at the beginning of the runs. Note that our use $r_{initial}$ is circumscribed only to the initial instability criterion.

\subsection{Initial conditions and code} \label{sec:inconds:code}
We have chosen a box with size $L_{b} = 70\, h^{-1}\, \rmn{Mpc}$,
not large enough to minimise cosmic variance effects. Within this box, we have performed two runs, differing only for the number of particles used:
run 32M was performed using $320^{3}$ particles, and run 500M with 
$800^{3}$. Thus, the particles masses for runs 32M and 500M are,
respectively, $7.49\times 10^{8}$ and $4.79\times 10^{7}\,$ $h^{-1}\,$
M$_{\sun}$.\\
\noindent
We have generated the initial conditions using a parallel version of the \textsc{IC} package by \citet{2005ApJ...634..728S}, that we developed ourselves. We started all the simulations from a redshift $z=50$. The latter is chosen in such a way to ensure that the linear modes are linear within the chosen box \citep[][sec. 4.1]{2007ApJ...671.1160L}. Recent work \citep{2009ApJ...698..266K} suggests however that, for the mass range of interest to this work, the starting redshift has little influence on the average final properties of the haloes.\\
\noindent
Our main simulation tool is \textsc{FLY}, a parallel MPI N-body cosmological simulation code implementing a parallel version of the Barnes-Hut octal tree algorithm \citep{2003CoPhC.155..159A,2007CoPhC.176..211B}. \textsc{FLY} adopts an efficient parallelization scheme: domain decomposition is applied to distribute particles, and workload decomposition is further applied to ensure load balancing \citep[see][for a detailed description of the parallel algorithm implemented in \textsc{FLY}]{2000JCoPh.163..118B}.

\subsection{Extraction of haloes} \label{sec:haloes:extr}
The availability of large samples of DM haloes, obtained from state-of-the art N-body simulations, means we can obtain statistics such as the mass function with very small poissonian errors. However, in order to extract a significant amount of cosmological information from these statistics, one should have a control on the systematic effects. The \emph{definition} of DM halo is one possible source of systematics. Presently, halo definition algorithms fall into two broad categories: those based on the \emph{friends of friends} (FOF) algorithm, and those based on some \emph{threshold overdensity} criterion, both of which are usually supplemented by some recursive scheme to eliminate outliers (i.e. gravitationally unbound particles). Different definitions mostly affect small haloes and the outer, low-density regions of more massive haloes \citep{2007ApJ...671.1160L}. For this reason, we have chosen to restrict our attention to haloes described by a relatively large (300) minimum number of particles.\\
\noindent
We have decided to adopt for our analysis a group finder which has only one threshold parameter, i.e. the recently introduced \textsc{AMIGA Halo Finder} \citep{2004MNRAS.351..399G,2009ApJS..182..608K}.
In \textsc{AHF} haloes are defined using the commonly used virialization criterion, i.e. first computing isodensity contours, and then including only those particles whose average density is larger than the critical oversdensity:
\be
\bar{\rho}_{c}(r_{v})/\rho_{b} = \Delta_{vir}(z)
\label{eq:sim:he:1}
\ee
{In the above equation $\bar{\rho}_{c}$ is the average density within the virial radius of each halo, $\rho_{b}$ the background density, and} the virial overdensity $\Delta_{vir}(z)$ depends on the cosmological model. This is not the only possible physically meaningful definition of the mass of a numerical halo \citep[see][for a discussion]{2001A&A...367...27W}: the most relevant differences in these definitions arise in the identification of the outer boundary of a halo, which can in principle have a significant impact also on the derived \emph{statistical} properties. As we stated above, we minimize discreteness effects by including only haloes having a number of particles $N_{p} \ge 300$, thus restricting the mass range to $M \ga 2.247\times 10^{11}$ and $1.437\times 10^{10}\, \rm{M}_{\odot}$, for the 32M and 500M runs, respectively.\\
\noindent
{In {\sc AHF} the gravitational energy of each particle is computed using the same tree data structure used in the octal tree codes to compute the gravitational force. Thus, the potential energy of haloes is computed without making any hypothesis concerning the shape or density distribution of the haloes themselves.} 

\section{Angular momentum distribution}  \label{sec:evol:props}
Due to the rather limited size of the box we have employed, cosmic variance could have an influence on the statistical properties of the halo sample. In this section we will then first investigate the angular momentum--mass relation.  
 
\subsection{Angular momentum--mass relation}  \label{sec:jm}
One of the most important predictions of tidal torque theory is that the angular momentum--mass relationship of relaxed haloes should be described by a power law: $J\propto M^{\alpha}$, with $\alpha=5/3$ \citep{1969ApJ...155..393P}. 
\noindent
\begin{figure}
\centering
\includegraphics[scale=0.35,angle=0]{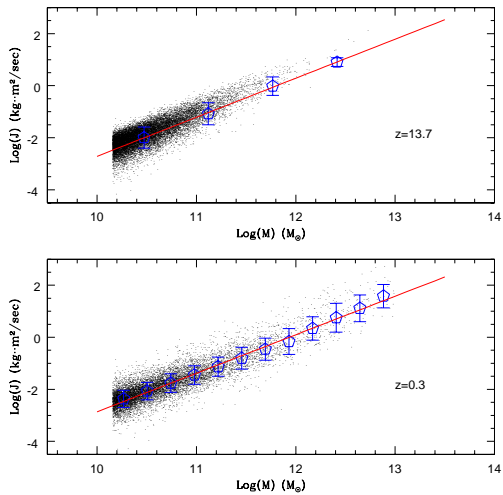}
\caption{Evolution of the $J-M$ relationship. The continuous lines are the best fits for two redshifts: $z=13.7$ (\emph{upper}) and $z=0.3$ (\emph{lower}), and the open symbols represent averages for the binned intervals. {Error bars are poissonian.} The mass of the haloes is evaluated as $M_{vir}$.}
\label{fig_j_m}
\end{figure}
\noindent
We test this prediction on the 500M run, shown in Figure~\ref{fig_j_m}. We notice that the distribution has an apparent excess of haloes having an angular momentum larger than the best-fit value. This is more evident at $z=13.7$, where the fraction of unrelaxed, high-spin haloes is larger, but we notice that also at $z=0.3$ the trend seems to persist.\\
\noindent
The best-fit values for the power law coefficients are shown in Figure~\ref{fig_alpha_j_m}, and we notice immediately that they are statistically compatible with the theoretical value, although the value at $z=0.3$ is smaller (at the 1$\sigma$ level). There is then a deviation from linear theory predictions at late times, when structure growth was non--linear. One possible explanation could be that linear theory does not take into account mergers, and their role in determining the statistical evolution of haloes \citep{2002ApJ...581..799V}. {In the same Figure we also show the effect of changing the definition of halo mass on the fitted properties of the $J-M$ relationship. While in the lower part, and in the rest of the paper, we use the virial mass as deduced by \textsl{AHF}, in the upper part of this figure we show the effect of using a different definition of halo mass, recently introduced by Cuesta et al. (2008). As one can appreciate, the effect is very little.}

\begin{figure}
\centering
\includegraphics[scale=0.35,angle=0]{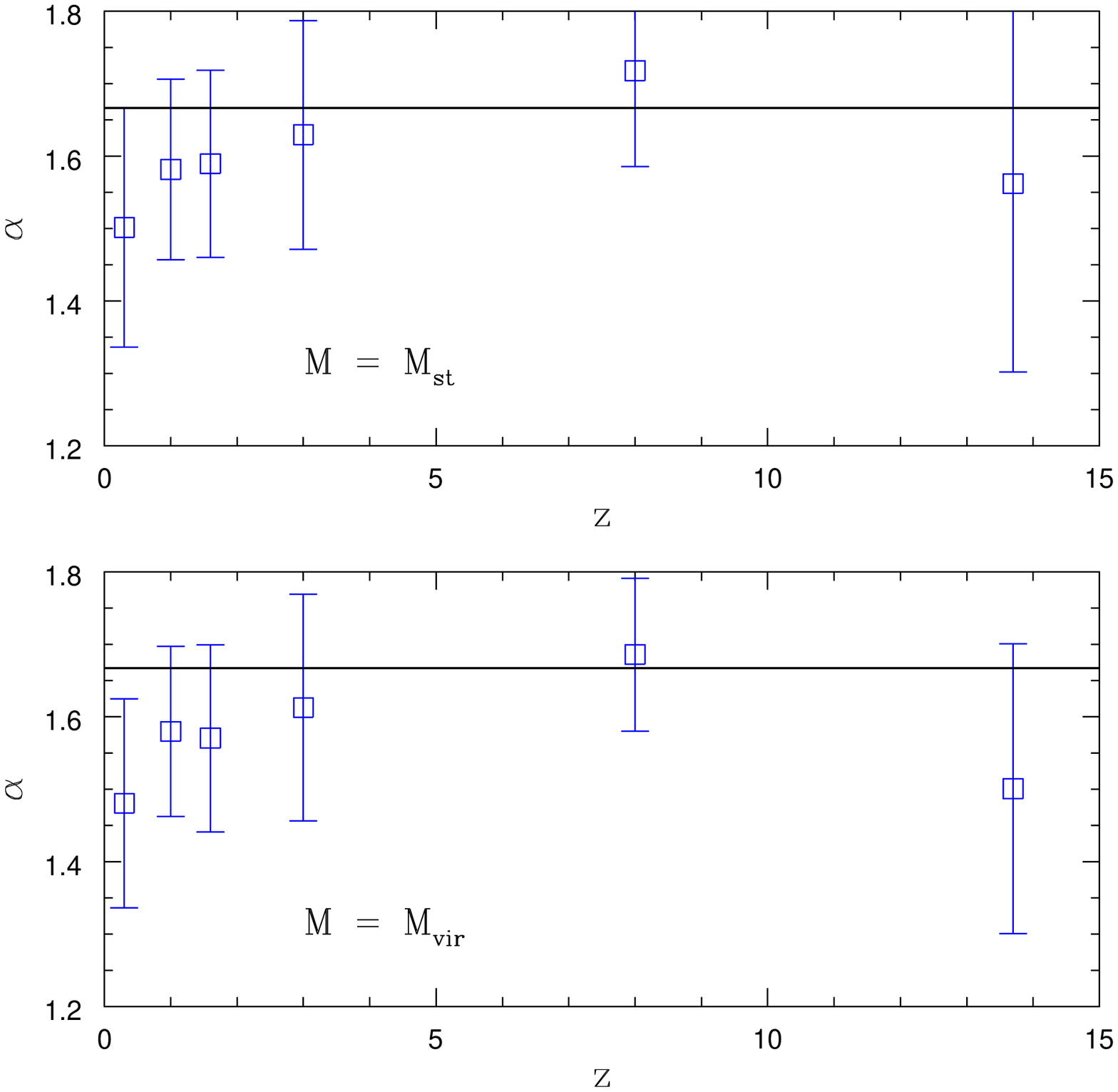}
\caption{Best fitted exponents of the J-M relationship. The upper plot shows results obtained using the \emph{static} halo definition introduced by \citet{2008MNRAS.389..385C}, the lower one is for the standard \textsl{AHF} halo mass definition. The horizontal continuous line show the theoretical $\alpha=5/3$ value from linear theory \citep{1969ApJ...155..393P, 1996MNRAS.282..436C, 1996MNRAS.282..455C}. {Error bars show the standard deviations in the values of the fitted slopes.} Although the upper values are slightly larger, the differences are within the uncertainties.}
\label{fig_alpha_j_m}
\end{figure}

\section{Dissecting the spin PDF}  \label{sec:diss:spin}

\begin{figure}
\centering
\includegraphics[scale=0.45,angle=0]{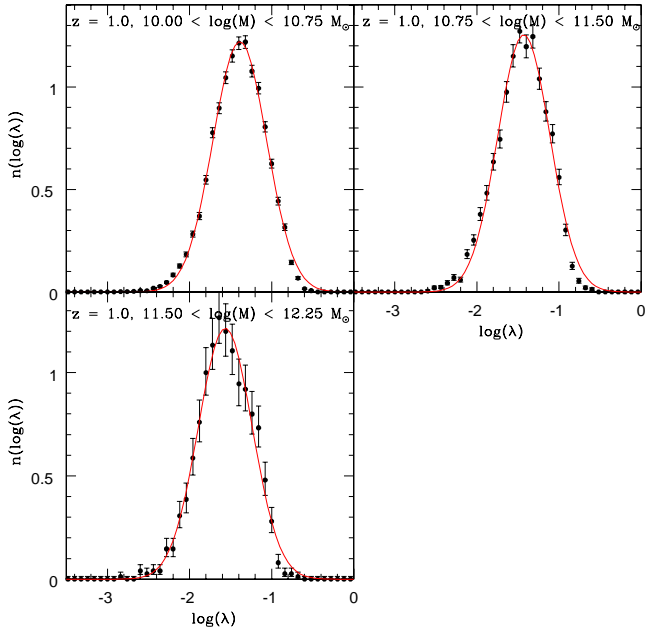}
\caption{Spin PDF at z=1, for different mass intervals. Error bars are Poissonian, computed according to the prescriptions in \citet{2003heinrich}. The curves are best-fitting lognormal functions (see Table~\ref{fit_lgn} for the values of $\bar{\lambda}$ and $\sigma_{{\rmn log}\lambda}$).} 
\label{fig:lam:mass:z1}
\end{figure}

\begin{figure}
\centering
\includegraphics[scale=0.45,angle=0]{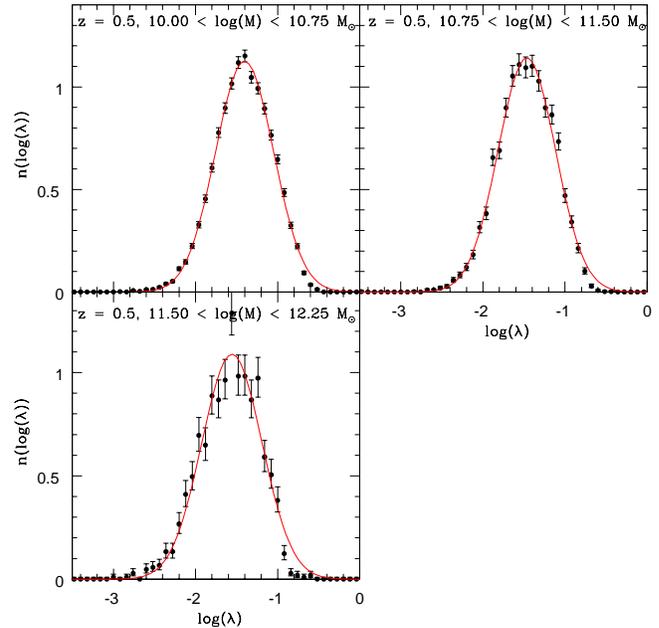}
\caption{Same as Fig.~\ref{fig:lam:mass:z1}, z=0.5. The PDF for the highest mass bin is not plotted, due to the poor statistics.} 
\label{fig:lam:mass:z0.5}
\end{figure}

\begin{figure}
\centering
\includegraphics[scale=0.45,angle=0]{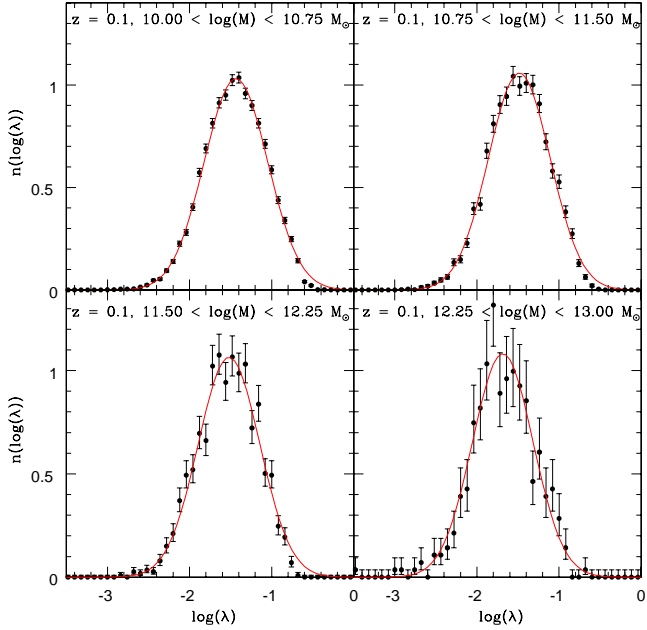}
\caption{Same as Fig.~\ref{fig:lam:mass:z1}, for redshift z=0.1.} 
\label{fig:lam:mass:z0.1}
\end{figure}

The evolution of the angular momentum distribution of dark matter haloes
has often been modelled using the \emph{tidal torque} theory
\citep{1951pca..conf..195H, 1955MNRAS.115....2S, 1969ApJ...155..393P, 1970Ap......6..320D, 1984ApJ...286...38W,2009IJMPD..18..173S},  
which predicts that a typical DM halo gains most of its angular momentum during the linear
regime \citep{1996MNRAS.282..436C}: it has further been shown that contributions from later collapse epochs and non-linear corrections are small \citep{2002ApJ...581..794C}.\\
\noindent
The spin parameter $\lambda$ (eq.~\ref{eq:am:0}) was introduced by
\citet{1971A&A....11..377P} as a convenient dimensionless quantity to
characterize the tidal growth of the angular momentum, also through the study of its \emph{probability distribution function} (PDF), $P(\lambda)$. All the quantities entering the definition of $\lambda$ depend ultimately on the adopted halo finder \citep{2007MNRAS.376..215B}.\\
\noindent
In order to understand the origin of this deviation, we will first consider the evolution of the spin PDF. In Figs~\ref{fig:lam:mass:z1}--~\ref{fig:lam:mass:z0.1} we show the evolution of the PDF for different mass intervals. We also plot the best-fit lognormal distributions, and in Table~\ref{fit_lgn} we present the parameters of these fits. Although the fitting functions of \citet{2007MNRAS.376..215B} and \citet{2008Ap&SS.315..191H} provide a better fit than a lognormal, the latter is a \emph{physically motivated} PDF distribution for the origin of angular momentum, predicted both by tidal torque and merger models. Thus, we will try to understand the possible physical origin of the \emph{deviations} from the lognormal distribution.
It is apparent from these figures that there is deficit of high-$\lambda$ haloes ($\log{\lambda/{\bar{\lambda}}} \ga 1.3 \div 2\,\sigma_{{\rmn \log}\lambda}$), at all redshifts, as was observed previously \citep{2001ApJ...557..616G, 2006ApJ...646..815S, 2007MNRAS.376..215B, 2008Ap&SS.315..191H}. This feature does not significantly depend on the mass, and in the next section we will argue that it originates from statistical correlations between quantities entering the definition of spin, and from deviations from lognormal behaviour of the energy $E$.\\
\noindent
From Table~\ref{fit_lgn} we also notice that the {location of the maximum} $\bar{\lambda}$ depends on the mass, being a slowly decreasing function of mass. This fact has been previously noticed by {various authors} \citep{1996MNRAS.281..716C,2007MNRAS.378...55M,2007MNRAS.376..215B,2008ApJ...678..621K}, but they suggested that the $\bar{\lambda}-$M relationship flattens for $z\la 1$, while our simulations show instead that also at more recent epochs the relationship holds true. Our findings and those of Knebe and Power are different to those of \citet{1999MNRAS.302..111L}, but more consistent with those of \citet{2007MNRAS.378...55M}, who find a very mild decreasing trend.
 \begin{table}
\caption{Parameters of the lognormal fits of the PDFs shown in
  Fig.~\ref{fig:lam:mass:z0.1}-~\ref{fig:lam:mass:z1}. Columns are as follows: (1) Mass interval
  (in \msunh), (2) $\bar{\lambda}$ (3) $\sigma_{\log\lambda}$ (4) redshift }
\begin{center}
\begin{tabular}{lccr}
\hline
\hline
log(M) & $\bar{\lambda}$ & $\sigma_{\log\lambda}$ & z\\
\hline
10.00 -- 10.75 & $0.041$ & $0.327$ & 1\\
10.75 -- 11.50 & $0.037$ & $0.318$ & 1\\
11.50 -- 12.25 & $0.027$ & $0.329$ & 1\\

& & & \\
10.00 -- 10.75 & $0.039$ & $0.354$ & 0.5\\
10.75 -- 11.50 & $0.034$ & $0.349$ & 0.5\\
11.50 -- 12.25 & $0.028$ & $0.367$ & 0.5\\

& & & \\
10.00 -- 10.75 & $0.037$ & $0.387$ & 0.1\\
10.75 -- 11.50 & $0.033$ & $0.377$ & 0.1\\
11.50 -- 12.25 & $0.030$ & $0.375$ & 0.1\\
12.25 -- 13.00 & $0.021$ & $0.369$ & 0.1\\

\hline
\end{tabular}
\end{center}
\label{fit_lgn}
\end{table}
Finally, we notice an evolution also in the shape of the best-fitting lognormal function, in the sense that the dispersion $\sigma_{{\rmn log}\lambda}$ seems to increase with decreasing redshift (third column of Table~\ref{fit_lgn}). We do not have sufficient statistics for $z\ga 0.1$ to determine if in the highest mass bin the high-$\lambda$ deficit and the $\sigma_{{\rmn \log}\lambda}$ width follow the same trend.\\
\noindent
\begin{figure}
\centering
\includegraphics[scale=0.45,angle=0]{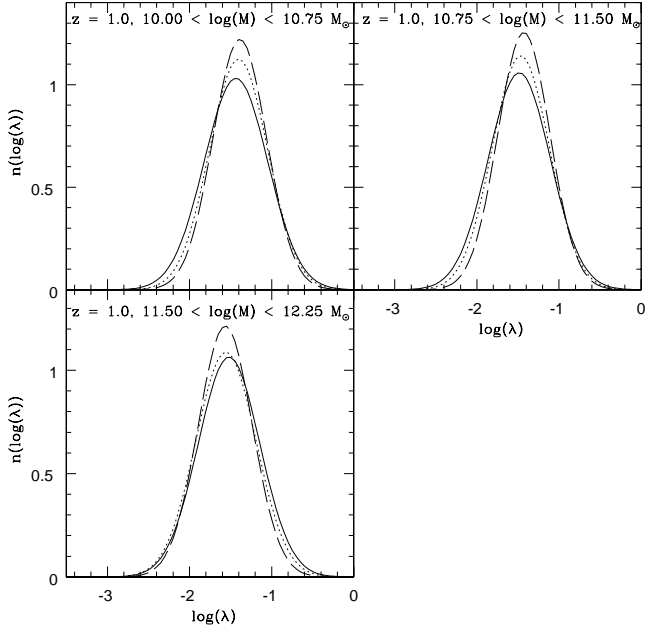}
\caption{Evolution of the lognormal fits to the spin PDF, for different mass bins. The different curves show the fits for different redshifts: \emph{continous}: $z=0.1$, \emph{dotted}: $z=0.5$, \emph{dashed}: $z=1$.} 
\label{evol:lgnfit}
\end{figure}
We summarize all these features in Figure~\ref{evol:lgnfit}, which shows an interesting fact: in the high mass bin ($10^{11.5}\le {\rm M} \le 10^{12.25}$) the {location of the maximum} $\bar{\lambda}$ stays almost constant after $z=1$.\\
\noindent
\begin{figure}
\centering
\includegraphics[scale=0.45,angle=0]{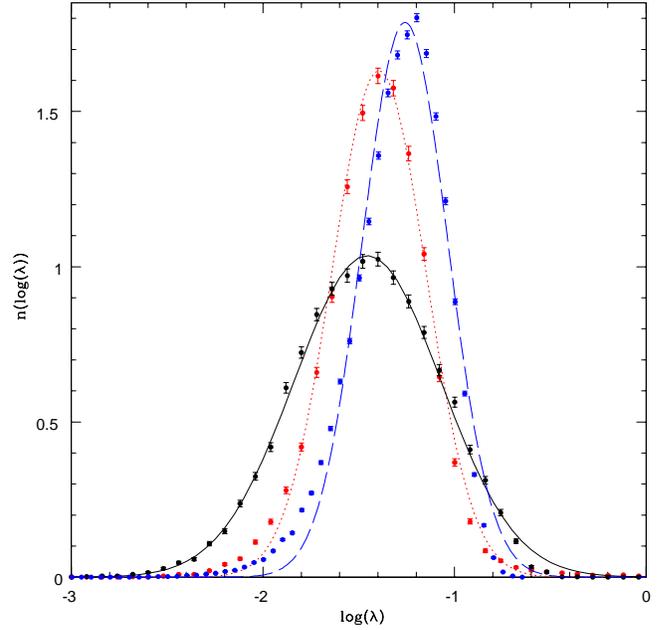}
\caption{Comparison of spin PDF at redshift z=0.1 for different simulations. The continuous curves are best-fitting lognormal functions for three different runs: run 500M (\emph{continuous, black}), \citet{2008MNRAS.391.1940M} (\emph{dotted, red}), and \citet{2008MNRAS.386.2135G} (\emph{dashed, blue}).} 
\label{fig:lam:compar:1}
\end{figure}
\noindent
\begin{figure}
\centering
\includegraphics[scale=0.45,angle=0]{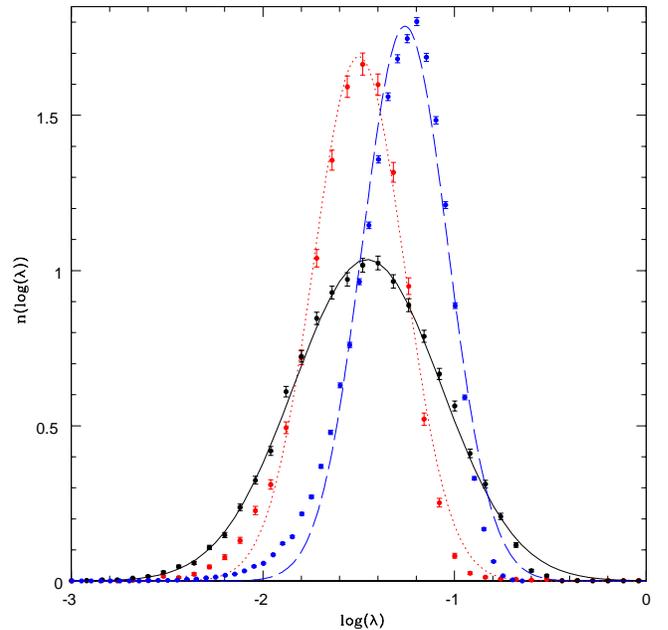}
\caption{Same as Fig.~\ref{fig:lam:compar:1}, but here the dotted red line uses only the relaxed haloes from the simulation of Macci\`{o} et al. are plotted.} 
\label{fig:lam:compar:2}
\end{figure}
Although these deviations from the lognormal fit are beyond statistical uncertainty, one could still suspect that they could be a product of some systematic effects, for instance introduced by the adopted halo finder. We have then checked the spin PDF distributions for two different simulations: the \textsc{GIF2} run \citep{2008MNRAS.386.2135G}, and a more recent simulation performed by \citet{2008MNRAS.391.1940M}. {We only use the halo catalogues provided by these simulations, instead of directly applying the \textsc{AHF} to their raw data. In the \textsc{GIF2} haloes were identified using a spherical overdensity criterion, without pruning the halo of gravitationally unbound particles. Also Macci\'{o} et al. use a spherical overdensity criterion, but they select haloes based on a number threshold criterion $N_{p} \geq 250$.}\\
While Giocoli et al. use the same definition of spin as we do (eq.~\ref{eq:am:0}), Macci\`{o} et al. use instead the defintion introduced by \citet{2001ApJ...555..240B}, and we convert it to ours using the prescirption described after eq. 4 of \citet{2007MNRAS.378...55M}.\\
\noindent 
In Fig.~\ref{fig:lam:compar:1} we show the PDFs for these simulations. The most striking feature of this figure is surely the great discrepancy in the distribution width (rms $\sigma_{\log\lambda}$) of run 500M wrt the other runs. Differences in the quantity $\bar{\lambda}$ are less pronounced, though still present. The values we find are however consistent with those previously found by other authors, as recently summarised in Fig.~7 of \citet{2006ApJ...646..815S} (see also their section 4.1). The reasons of such a result are still under investigation by us. Perhaps, the cause of these differences is to be found in the different ratio between the amount of accretion- or merger-dominated spin histories \citep[but see][]{2007MNRAS.380L..58D}. Another interesting feature is that the PDFs of both Giocoli et al. and Macci\`{o} et al. show an excess of low spin haloes, and a deficit in the high-$\lambda$ tail with respect to a lognormal, while the PDF for the 500M run only shows the latter feature. {A similar excess of low-spin haloes was observed e.g. by \citet{2007MNRAS.376..215B}, and seems to be a general feature of many other simulations. Note however that the mass range of our simulations is more biased towards galaxy-sized haloes, while low-spin haloes tend to be those with higher mass \citep{2008ApJ...678..621K}. The excess of low-spin haloes seems to be increasing with the average mass range probed by the simulation, which is larger in the \textsc{GIF2} simulation, and smaller in our \textsc{run 500M}. On average, low-spin, high-mass haloes could be less relaxed at any given epoch: this could explain why our haloes seem to better follow a lognormal distribution.}\\
\noindent
\citet{2008MNRAS.391.1940M} introduce a sophisticated criterion to isolate \emph{relaxed} haloes (see section 2.2. of their paper): however, we see from Fig.~\ref{fig:lam:compar:2} that these relaxed haloes have a spin PDF showing the same qualitative features as the global one. {Also, we note that \citet{2007MNRAS.376..215B} found that picking just relaxed haloes made a big difference on
the form of the spin distribution, but that the biggest difference was
due to the halo-finder rather than imposing a cut on the instantaneous
"virial ratio". Thus, the difference between relaxed and unrelaxed haloes could be very slight, in the light of these differences.}\\
\noindent
We then conclude that the spin PDF at all redshifts since $z\sim 0.5$ shows a detectable deficit of haloes having $\log{\lambda/{\bar{\lambda}}} \ga 1.3 \, \sigma_{{\rmn \log}\lambda}$, with respect to a lognormal distribution. This deficit is quite detectable and statistically robust, and apparently is a feature not depending on the particular halo finder or dynamical state of the halo.\\
\noindent 
In the next subsection we will use a very general statistical theorem to understand the origin of this feature in the spin PDF.
\subsection{Statistical origin of deviations}
\begin{figure}
\centering
\includegraphics[scale=0.35,angle=270]{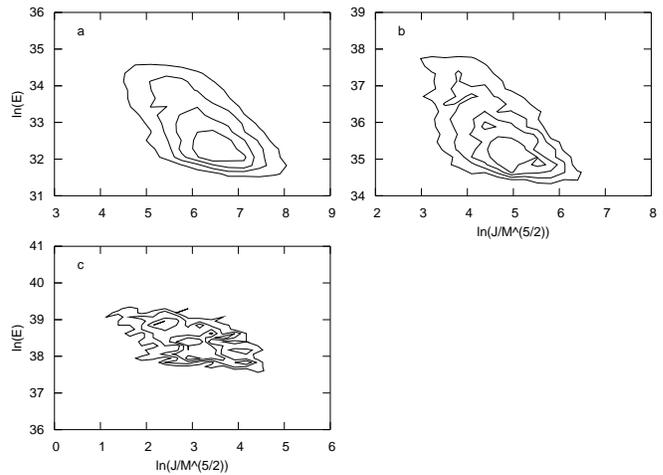}
\caption{Correlation diagrams for the quantities in the right-hand
  side of eqn.~\ref{eq:pdf:1}, at redshift z$=1$. {Here and in Figures 10--12 axes units are arbitrary.} The four levels are density contours as fractions of the maximum, at the $80\%, 60\%, 40\%$ and $20\%$, levels, from the inner- to the outermost. The three plots are for different mass intervals, corresponding to those of the PDF distribution of Fig.~\ref{fig:lam:mass:z0.1}: \emph{upper left}: $10 \le {\rmn \log}({\rm M}) < 10.75$, \emph{upper right}: $10.75 < {\rmn \log}({\rm M}) \le  11.50$, \emph{lower left}: $11.50 < {\rmn \log}({\rm M}) \le 12.25$. }
\label{fig:corr:dstr1}
\end{figure}

\begin{figure}
\centering
\includegraphics[scale=0.35,angle=270]{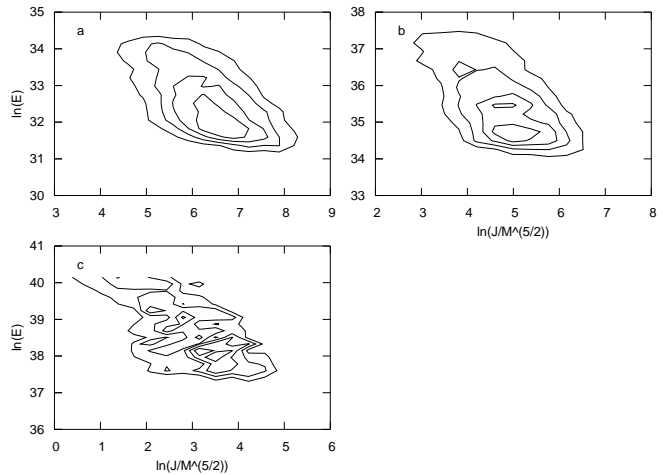}
\caption{Same as Fig.~\ref{fig:corr:dstr1}, for z=0.5.}
\label{fig:corr:dstr0.5}
\end{figure}

\begin{figure}
\centering
\includegraphics[scale=0.35,angle=270]{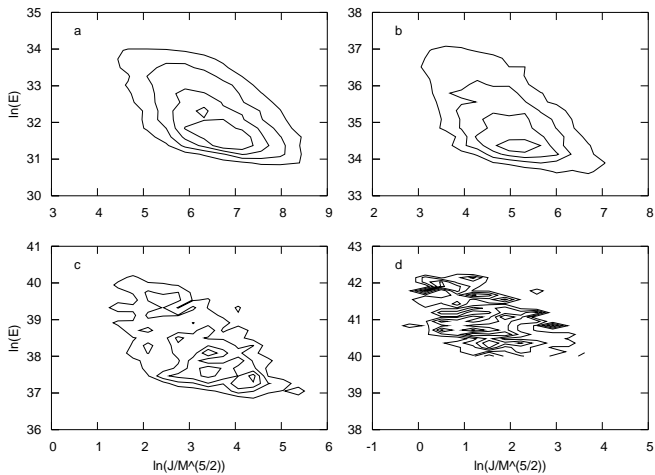}
\caption{Same as Fig.~\ref{fig:corr:dstr1}, for z=0.1. In addition to the three mass intervals, in the , lower right section we plot the contour plots for the highest mass bin: $12.25 < {\rmn \log}({\rm M}) \le 13$. }
\label{fig:corr:dstr0.1}
\end{figure}

The spin $\lambda$ can be regarded as a stochastic variable depending on three other variables: $J, E\, \rmn{and}\, M$. We can decompose the spin in two terms, plus a constant:
\be
{\rmn \ln}(\lambda) = {\rmn \ln}\left(\frac{J}{M^{5/2}}\right) + \frac{1}{2}{\rmn \ln}(E) -{\rmn \ln}(G)
\label{eq:pdf:1}
\ee
The first factor in the right-hand side can be regarded as depending on the mass $M$, in the light of the scaling $J\propto
M^{5/3}$, while the total energy $E$ also depends on the dynamical
state of the halo. The latter is also determined by environment, through e.g. tidal fields and streaming motions. We expect these two terms to be
distributed similarly for fully virialised haloes, {although the environmental dependence of $E$ could introduce an additional dependence in addition to that on the mass ${\rm M}$. Here however we only consider the distribution of the \textit{global} spin, without analysing environmental differences, but the reader should be made aware of the fact that environmental differences are not only modulated by the mass dependence, as recently shown by \citet{2010ApJ...708..469F}}.\\
Following Cram\'{e}r's theorem \citep{Cramer..MatZeit..41..405..1936}, two necessary and sufficient conditions for the variable ${\rmn \ln}(\lambda)$
to be normally distributed are that the two quantities in the
right-hand side of eq.~\ref{eq:pdf:1} are \emph{statistically
  independent} and normally distributed. Thus, we expect that 
deviations from a lognormal fit arise if at least one the two terms in eq.~\ref{eq:pdf:1} is not normally distributed, and/or if there are correlations between 
$J/M^{5/2}$ and $E$. In our case, \emph{both} situations are taking place.\\
In order to check the independence of the two terms, we show in
Figure~\ref{fig:corr:dstr1}-~\ref{fig:corr:dstr0.1} contour plots of the quantities $J/M^{5/2}$ and $E$. 
It is evident that the two quantities show correlations in all mass bins, although these become progressively uncorrelated with increasing mass. The lack of correlation in the larger mass bin (bottom right plot in Fig.~\ref{fig:corr:dstr0.1}) is probably partly due to the lack of very large haloes in our sample, a consequence of cosmic variance. 
\begin{table}
\caption{Kendall's correlation indexes t$_{\rmn k}$ between $J/M^{5/2}$
  and $E$. A value of the correlation index lying in the
  interval $[-0.5, 0.5]$ is taken as evidence of poor correlation.}
\begin{center}
\begin{tabular}{lcr}
\hline
\hline
${\rmn \log}$M ([M$_{\odot}$]) & t$_{\rmn k}$ & z\\
\hline
10.00 -- 10.75 & -0.391 & 1\\
10.75 -- 11.50 & -0.400 & 1\\
11.50 -- 12.25 & -0.329 & 1\\
& & \\
10.00 -- 10.75 & -0.382 & 0.5\\
10.75 -- 11.50 & -0.381 & 0.5\\
11.50 -- 12.25 & -0.427 & 0.5\\
& & \\
10.00 -- 10.75 & -0.368 & 0.1\\
10.75 -- 11.50 & -0.388 & 0.1\\
11.50 -- 12.25 & -0.391 & 0.1\\
12.25 -- 13.00 & -0.297 & 0.1\\
\hline
\end{tabular}
\end{center}
\label{kendall_fit}
\end{table}
The outputs of Kendall's correlation test \citep{1979ats..book.....K} are given in Table~\ref{kendall_fit}. They show that the two quantities become
progressively more statistically independent with increasing mass, and that the correlations are actually never very significant, in statistical terms. This is also evident from the plots, which show that the two outermost contours ($40 \,{\rm and}\, 20\%$ of the maximum), are partially elongated. Although small, this  correlation can partially explain the presence of the deviations of the spin from a lognormal distribution.\\
\begin{figure}
\centering
\includegraphics[scale=0.45,angle=0]{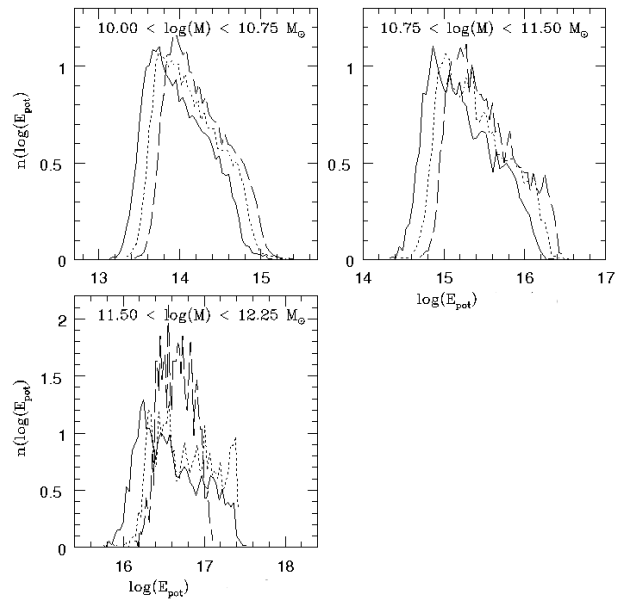}
\caption{PDF of the total energy $E$, for z$=1.$ (continuous), z$=0.5$ (dotted), z$=0.1$ (dashed). We do not show the PDF for the highest mass bin, which is noise dominated.}
\label{fig:pdf:egrav}
\end{figure}
\noindent
Finally, we have checked whether one or both of the two quanties entering our decomposition of the spin $\lambda$, i.e. $J/M^{5/2}$ or $E$, are lognormally distributed. In Figure~\ref{fig:pdf:egrav} we show the PDF of ${\rmn \log}(E)$, which shows a shoulder at relatively large values, at least up to $M \la 3.6\times 10^{11}$ M$_{\odot}$. For larger masses, the distribution deviates even more, a sign of poor relaxation of these haloes. Note that the dynamical relaxation of haloes generally takes longer, affecting the gravitational binding energy and thus the total energy, so that this could explain the lack of a lognormal distribution of the latter.

\section{Conclusions}   \label{sec:concl}
The work we have presented in this paper aimed at understanding the origin of the \emph{deviations} of the spin PDF from a lognormal distribution. Such deviations have been recently noticed in numerical simulations, and even predicted by statistical models for the angular momentum growth \citep{2002ApJ...581..794C}. Instead of attempting some \emph{ad-hoc} fitting function for the PDF, we have attempted to understand the statistical origin of these deviations. To that end, we have perfomed two high-resolution simulations (maximum resolution: $m_{\rmn p} = 4.37\times 10^{7}$ M$_{\odot}$), in a box of $L_{b} = 70 h^{-1}$ Mpc, which provided us with a large number ($\sim 16,000$) of well resolved haloes, i.e. haloes having more than 300 particles. Thus, we believe that our results are statistically robust.\\
\noindent
We can summarise our main findings in the following points:
\begin{itemize}
\item The J-M relation is well fitted by a power law, with exponent $\alpha = 5/3$, except at very recent epochs;
\item The spin PDF is systematically low at $\log{\lambda/{\bar{\lambda}}} \ga 1.3 \, \sigma_{{\rmn \log}\lambda}$.
\end{itemize}
We have concentrated our attention on intepreting the latter point, using basic statistical arguments (Cram\'{e}r's theorem). An implicit assumption in our approach is that the lognormal distribution should be in some way preferred over other possible functional forms. Notice that not only the ''nearest neighbours'' tidal torque theory predicts such an output \citep{1996MNRAS.282..436C,1996MNRAS.282..455C}, but also merging-based models like those of \citet{2002ApJ...581..799V} and, more recently, the extended Press-Schechter models by \citet{2008Ap&SS.315..191H}. Some of these models were based on the analysis of N-body simulation outputs, with a mass resolution down to $m_{\rmn p} \sim 10^{7}$ M$_{\odot}$, but using a set of nested boxes \citep[e.g][]{2007MNRAS.378...55M}. In order to probe halo formation using a comparable mass resolution, while avoiding the technical problems arising from the correct reproduction of tidal fields, that arise when using nested boxes, we have performed a larger simulation, and analysed the evolution of the mass function.\\ 
{\citet{2007MNRAS.376..215B} have provided arguments to support the idea that a PDF which is not anti-biased against small ($\lambda\simeq 0$) values of the spin (a typical feature of the lognormal distribution) should provide a better description of the actual spin PDF. However, our results do not seem to support their conclusion: particularly at low values of $\lambda$ the lognormal seems to provide a very good fit of our simulations. One possible reason of this discrepancy could be due to the lies in the fact that our simulations, compared to the \textsc{Millennium} run analysed by Bett et al., do not have haloes more massive than $\approx 10^{12} {\rm M}_{\sun}$, while in the \textsc{Millennium} run this upper limit extends to ${\rm M}\approx 10^{15} {\rm M}_{\sun}$. If high mass haloes tend to have a smaller spin \citep{2008ApJ...678..621K}, then we conclude that our simulations probe better the mass range typical of low- to intermediate mass galactic haloes, avoiding instead massively galactic and cluster-sized haloes. The latter tend to be not completely relaxed, thus their spin could deviate from the lognormal form expected from Tidal Torque Theory.}\\
We have chosen a relatively small cosmological volume ($L_{b} = 70\, h^{-1}\, {\rm Mpc}$), in order to test whether the angular momentum--mass relation and spin distributions show \emph{statistically detectable} deviations from the predicted analytical forms they have for dynamically relaxed haloes. The large statistics made possible by the high resolution allows us to detect some significant deviations: however, any numerical experiment is subject to some systematics.\\
\noindent
We cannot rule out that some residual systematic uncertainties could arise from the adopted halo finder: the actual spin of a numerical halo should also depend on the adopted halo finder, i.e. on the numerical definition. Even in the ''nearest neighbours'' tidal torque theory by \citet{1969ApJ...155..393P} the distribution of angular momentum within the (spherically averaged) collapsed halo reflects the evolution of the torque during the collapse: the torques on shells placed at larger distances decrease on average with radius. Thus, one could expect that the distribution of $\lambda$ would depend on the radius at which it is estimated, and this is generally different for halo finders based on critical isodensity contours or on friends-of-friends algorithms.\\
\noindent
{Concerning the environmental dependence of the spin distribution, it is now evident that the original claim by \citet{1999MNRAS.302..111L}, who} did not find any evidence for this dependence, {was due to a lack of sufficent numerical resolution in their simulations. Already} in a previous work \citep{2002MNRAS.332....7A} we found that the spin's PDF has a significantly smaller dispersion and mean in Voids, a result consistent with the more recent findings that a higher merger frequency tends to produce haloes with higher spin \citep{2001ApJ...557..616G,2002MNRAS.329..423M,2006MNRAS.370.1905H}. \\
Our simulations have sufficient statistics to test the predictions of the linear model concerning the $J$--$M$ relationship. We notice that the index $\alpha$ of the power law $J\propto M^{\alpha}$ at redshift zero is smaller than the standard value $5/3$. A similar fact had been previously noticed by \citet[][see their Table 6]{2000MNRAS.311..762S}, using simulations on a larger volume, thus it is not probably a peculiarity of our use of a small volume.\\
\noindent
We measure average values of the spin at $z\sim 0.1$ consistent with those found recently by other authors \citep{2007MNRAS.378...55M}: $\bar{\lambda}\simeq 0.03$. Macci\`{o} et al. find little dependence of  $\bar{\lambda}$ on halo mass, even when they restrict their analysis to relaxed haloes. Our results are consistent with their findings, and with theoretical analyses \citep{2002ApJ...581..794C}. We do however see a dependence of the \emph{spin dispersion} on mass: more massive haloes tend to have a lower dispersion, although it is a very slight trend which decreases towards low redshifts, consistently with the idea that it is induced by the presence of not fully relaxed haloes.  
\\
\noindent
The deviations of the spin's PDF from a lognormal arise because of statistical correlations between mass and angular momentum {and because of significant deviations of (at least) the total energy $E$ from a lognormal distribution}, as implied by Cram\'{e}r's theorem. These correlations tend to disappear when haloes collapse, as one should expect for {ideal} completely relaxed haloes$^1$\footnote{For instance, in the spherical collapse model the final overdensity only depends on the value of the matter overdensity $\Omega_{m}$  at collapse redshift \citep{1972ApJ...176....1G}}. Note that this conclusion is true both for merger- {and} collapse-induced relaxation: {in either case, the result is that the halo retains no "memory" of its previous state, due to dynamical friction acting on dynamical timescales.}

\section*{Acknowledgments}
V. A.-D. gratefully acknowledges the hospitality of the Institute for Theoretical Astrophysics of the Ruprecht-Karls University of Heidelberg, where part of this program was performed.\\
\noindent
The work of V. A.-D. has been supported by grant no. 330 of the HPC-Europa 2 Transational Access Program. A.D. acknowledges the Slovak Academy of Sciences Grant No. 2/0078/10 and HPC Europa grants HPC04MXW87 and HPC0477ZZL. A.D.R. acknowledges partial support from ALMA-CONICYT FUNDS through grant
32070013 and 31070023, from FONDECYT - Proyecto de Iniciaci\'{o}n a la Investigaci\'{o}n No. 11090389, and from UNAB - Proyecto Regular No. DI-35-09/R.\\
\noindent
This work makes usage of results produced by the PI2S2 Project managed
by the Consorzio COMETA, a project co-funded by the Italian Ministry
of University and Research (MIUR) within the {\em Piano Operativo Nazionale 
"Ricerca Scientifica, Sviluppo Tecnologico, Alta Formazione" (PON
2000-2006)}. More information is available at http://www.pi2s2.it (in
italian) and http://www.trigrid.it/pbeng/engindex.php .\\
\noindent
{Finally we would like to acknowledge the referee for his/her comments, which greatly improved the paper.}


%
\bibliographystyle{/home/antonucc/tex/mnras_tex2e/mn2e}
\bibliography{/home/antonucc/papers/ref_j/biblio_nb}


\label{lastpage}

\end{document}